# #Change: How Social Media is Accelerating STEM Inclusion

*Authors:*

*2030STEM Salon Series. 2030STEM Inc., NY, NY USA*

*Jennifer D. Adams, PhD, 2030STEM Salon Series Editor, University of Calgary, Calgary, AB, CAN,*

*Carlotta A. Berry, PhD ROSE-HULMAN INSTITUTE OF TECHNOLOGY, 5500 Wabash Avenue, Terre Haute, IN, USA*

*Ruth Cohen, Interim Executive Director and Strategic Advisor, 2030STEM Inc, NY, NY, USA*

*Alonso Delgado, Dept. Evolution, Ecology and Organismal Biology, The Ohio State University, Columbus OH, USA*

*Jackie Faherty\*, PhD, CoFounder 2030STEM, The American Museum of Natural History, NY, NY USA*

*Eileen Gonzales, PhD, Department of Astronomy, Cornell Center for Astrophysics and Planetary Science, and Carl Sagan Institute, Cornell University, Ithaca, NY USA*

*Mandë Holford\*, PhD, CoFounder 2030STEM Hunter College, The American Museum of Natural History, CUNY Graduate Center, NY, NY USA*

*Ariangela J Kozik, PhD, Department of Internal Medicine, Division of Pulmonary and Critical Care Medicine, University of Michigan, Ann Arbor, MI, USA*

*Lydia Jennings, PhD, Community, Environment and Policy, College of Public Health, The University of Arizona, Tucson, AZ, USA*

*Alfred Mays, Director and Chief Strategist for Diversity and Education, Burroughs Wellcome Fund*

*Louis J. Muglia, MD PhD, Burroughs Wellcome Fund, Research Triangle Park, NC; Department of Pediatrics University of Cincinnati College of Medicine, Division of Human Genetics, Cincinnati Children's Hospital Medical Center, Cincinnati, OH, USA*

*Nikea Pittman, PhD, Department of Biochemistry and Biophysics, School of Medicine, University of North Carolina at Chapel Hill, Chapel Hill, NC, USA*

*Patricia Silveyra, PhD, Department of Environmental and Occupational Health, School of Public Health, Indiana University Bloomington, Bloomington, IN, USA*

*\*Co-Corresponding authors: [jfaherty@amnh.org](mailto:jfaherty@amnh.org) ; [mholford@hunter.cuny.edu](mailto:mholford@hunter.cuny.edu)*

## ABSTRACT

The vision of 2030STEM is to address systemic barriers in institutional structures and funding mechanisms required to achieve full inclusion in Science, Technology, Engineering, and Mathematics (STEM) and provide leadership opportunities for individuals from underrepresented populations across STEM sectors. 2030STEM takes a systems-level approach to create a community of practice that affirms diverse cultural identities in STEM.

This is the first in a series of white papers based on 2030STEM Salons—discussions that bring together visionary stakeholders in STEM to think about innovative ways to infuse justice, equity, diversity, and inclusion into the STEM ecosystem. Our salons identify solutions that come from those who have been most affected by systemic barriers in STEM.

Our first salon focused on the power of social media to accelerate inclusion and diversity efforts in STEM. Social media campaigns, such as the #XinSTEM initiatives, are powerful new strategies for accelerating change towards inclusion and leadership by underrepresented communities in STEM. This white paper highlights how #XinSTEM campaigns are redefining community, and provides recommendations for how scientific and funding institutions can improve the STEM ecosystem by supporting the #XinSTEM movement.

## OVERVIEW

The lack of full demographic representation in Science, Technology, Engineering, and Mathematics (STEM) has been a persistent challenge, rooted in historical and systemic exclusion of Black, Latino/a/x, and Indigenous people and other underrepresented groups [1-5]. STEM fields have faced critical barriers in recruiting and retaining Black, Latino/a/x, and Indigenous individuals and other people of color. Eliminating these barriers is not simple. There is significant work ahead to achieve parity and representation in the STEM workforce [6]. As aptly stated by Dr. Winston Morgan, Director of Impact and Innovation in the School of Health Sport and Bioscience, University of East London, "No Black Scientist has ever won a Nobel [Prize]—that's bad for science, and bad for society [7]." The presence and visibility of a cultural and gender diversity of scientists is important towards overcoming disparities and achieving equity in STEM [8].

However, Black, Latino/a/x, Indigenous and other historically underrepresented groups have not remained idle. In 2020, using collective action, people underrepresented in STEM took matters into their own hands—and online. Building on a burgeoning social media movement, several STEM affinity groups cultivated online communities that reflected their STEM fields, and equally important, their cultural STEM-identity using variants of the #XinSTEM hashtag.

Our inaugural 2030STEM Salon focused on these numerous, new #XinSTEM initiatives. We explored the impact of these grassroots efforts, outlined barriers to their success, and developed recommendations for creating sustainability for #XinSTEM movements to thrive and to accelerate change.

## WHAT IS THE #XINSTEM MOVEMENT?

"#XinSTEM" refers to a set of grassroots, social-media-based initiatives that foster inclusion, representation, and discussions of diversity across STEM fields. Many #XinSTEM initiatives are founded by graduate students or postdoctoral scholars, like Stephanie Page, who founded #BLACKandSTEM in 2014 while still a PhD student.

#XinSTEM groups have proliferated to meet a need that is not being met and to combat entrenched power and privilege in STEM institutions and workplaces. These movements allow early-career (and other) researchers from underrepresented groups to connect, build community, and foster mentorship and advocacy across vast networks in real time. Communities share experiences and strategies for a myriad of activities that allow them to:

- amplify their voices
- tell their unique stories
- share their research and groundbreaking publications
- find invaluable resources
- access much-needed professional development
- leverage the power of a formal network for greater collaboration
- reach a larger academic and public community

Between 2019 and 2020, #XinSTEM social media grew exponentially. Figure 1 plots the meteoric rise of BlackinX entities as an example of the XinSTEM growth spurt. This unprecedented level of engagement and support created viral advocacy that has shed light on the discrepancies in STEM inclusion and leadership.

For the first time ever, in a virtual global community, underrepresented students, researchers, professors, and others working in STEM have fostered coordinated, networked, public discussions to accelerate the pace of change in their communities [9-10].

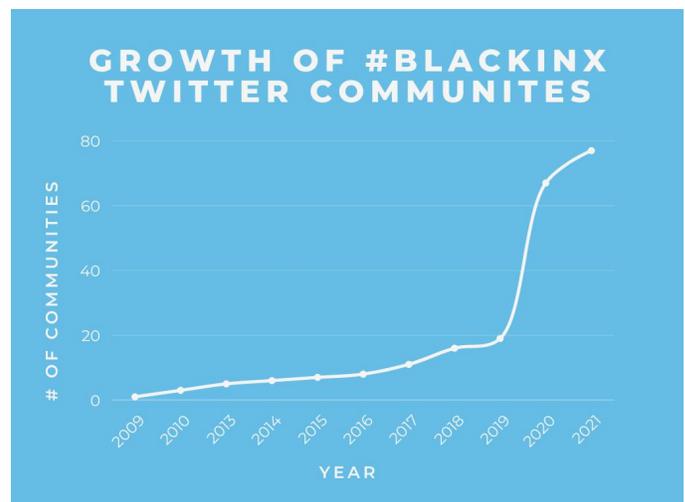

**Figure 1: The rise of BlackinX.** Cumulative plot outlines the rapid growth of BlackinSTEM movements from 2009-2021. Data provided by the BlackinX community collated by Carlotta Berry.

### #XinSTEM is a Catalyst for Change

The use of social media as a tool to catalyze social and institutional change is relatively recent. In her 2018 seminal study, Beronda L. Montgomery wrote that the impacts of #XinSTEM social media start-ups "can range from community building, to proactive mentoring and advocacy, as well as more customary uses for supporting scholarly success of diverse individuals, including dissemination and accessible discussions of research findings [9]."

#XinSTEM movements, in particular, have been recognized as creating powerful online communities via social media that can potentially catalyze lasting change in STEM fields.

At our 2030STEM Salon, #XinSTEM founders described the motivations, goals, and impacts of their movements. Many of these campaigns and collaborations have since moved beyond social media, transforming into organizations that plan and hold in-person networking events and professional development workshops. The following is a snapshot of some prominent #XinSTEM initiatives and their activities and approaches.

*NOTE: This is not an exhaustive list of #XinSTEM movements or collaborators.*

**#BlackinX.** Dr. Carlotta Berry, professor and Lawrence J. Giacoletto Chair of Electrical and Computer Engineering at Rose-Hulman Institute of Technology, is the co-founder of #BlackInRobotics and #BlackInEngineering. In 2020, she worked with several members of the #BlackInX community, including Samantha Mensah of #BlackInChem and Quincy Brown of #BlackInComputing. Together, they organized a network of over 80 groups working to advance inclusion and representation to host the inaugural #BlackInX conference, which took place from June 29 to July 3, 2021 [11]. It was considered a sort of "homecoming," marking the one-year anniversary of the exponential growth of the #BlackInX movement.

Both the turnout and engagement were significantly higher than expected. As Berry noted, "we were almost a little blindsided by the amount of people who wanted to connect with us. It was apparent that members of our community were hungry to connect with like-minded individuals to promote a shared mission."

**#BlackinPhysics.** Dr. Eileen Gonzales, a postdoctoral fellow at Cornell, along with Dr. Charles Brown, Dr. Jessica Esquivel, and 9 other early career researchers co-founded #BlackinPhysics to "celebrate Black Physicists and our contributions to the scientific community and to reveal a more complete the picture of what physics looks like." Within a short time, this hashtag amassed >3,000 followers and 1.3 million impressions on Twitter. In the last year the #BlackInPhysics movement has transformed into a non-profit organization Black In Physics lead by Drs. Gonzales, Brown, and Esquivel. During the past three year the organization has held professional and social events for the community during #BlackInPhysics Week and beyond. Some of these include: two Wiki-thons in partnership with the American Physical Society and the American Institute of Physics to update the Wikipedia pages of Black scientists, three job fairs showcasing jobs and internship opportunities in their communications, and a yearly essay series during #BlackInPhysics Week were published in *Physics Today* [12-14].

**#BlackInAstro**. Ashley Walker, an astrochemist and planetary scientist from Chicago, founded #BlackinAstro to dispel assumptions about scientists who are women, Black, or both. Emphasizing her African American Vernacular English (AAVE), she has noted that people often assume she's "less than," "subpar," "angry," and "ghetto." #BlackInAstro week has allowed Ashley to highlight astronomers of color, which she is hoping can help combat and dispel discriminatory attitudes in planetary science.

Through her leadership and that of others, #BlackinAstro has attracted funding from Merck and the Royal Society of Chemistry, has been supported by Ohio State University and UCLA, and has even attracted the attention and support of celebrities like Michael B. Jordan.

**#BlackinMicro.** Dr. Kishana Taylor, a postdoctoral fellow at Carnegie Mellon University, and Dr. Ariangela Kozik, a postdoctoral fellow at the University of Michigan, co-founded #BlackinMicro amidst a global pandemic with a disproportionate impact on communities of color. In 2020, the #BlackinMicro Week virtual conference drew 2,500 participants. By 2021, their conference attendees represented 76 different countries. They have over 9,000 Twitter followers, and posts during #BlackInMicro Week gained 1.1 million impressions.

Having highlighted a critical gap in the microbial sciences, #BlackInMicro has transitioned to a non-profit organization, the Black Microbiologists Association (BMA), to continue its work. BMA was founded by Drs. Taylor and Kozik, along with #BlackInMicro Week organizers Dr. Nikea Pittman, a postdoctoral fellow at the University of North Carolina at Chapel Hill; Dr. Chelsey Spriggs, a postdoctoral fellow at the University of Michigan; and Dr. Ninecia Scott, a postdoctoral fellow at the University of Alabama at Birmingham [15]. BMA currently consists of over 300 members and is working to secure long-term funding for programs and initiatives to further support Black microbiologists across career stages.

**#BlackInNeuro.** Angeline Dukes, a graduate student researcher/addiction neuroscientist from Warner Robins, Georgia, created #BlackinNeuro in response to growing up seeing only white men as scientists. #BlackInNeuro Week was an effort to highlight neuroscientists of color and create awareness in the field [16]. Since then, the movement has grown into a non-profit organization, hosting over 56 events and building a directory of 300+ Black neuroscientist profiles.

As a first-generation American and college student whose parents immigrated from Trinidad and Haiti, Dukes didn't get much guidance when it came to applying to schools, seeking scholarships, and choosing careers. She wants to make sure other Black students get the mentorship they need when it comes to choosing a career in science and pursuing their studies. Today, this vision connects scientists around the world, as #BlackInNeuro has amassed over 26,500 followers on Twitter.

**#BlackinNHM**. Adania Fleming, a naturalist at the Florida Museum of Natural History at the University of Florida, founded #BlackinNHM in February of 2021 because "Blacks need to build their own museum space." The narrow white culture of museums fails to represent her own cultural and international background.

The first #BlackinNHM week was held not only for Black naturalists, but also for Black museum-goers. The vision and aim of #BlackinNHM is a larger cultural change for all participants in natural history.

**#LatinXinMarineSciences.** A group of early-stage Ph.D. Students: Alonso Delgado (Ohio State University), Ivan Moreno (UC San Diego), Arani Cuevas (Texas A&M) and Alejandro De Santiago (University of Georgia) founded #LatinXinMarineSciences in the spring of 2020.

Marine science remains one of the least diverse fields of research, and the non-white students navigating it need support from a community of peers and teachers they can identify with. The goal of this organization was to find other Latin-American and Caribbean marine scientists and create a virtual community that fostered support, shared resources, and helped to actively diversify marine science.

**#LatinXinBME.** Dr. Brian Aguado (UC San Diego) and Dr. Ana Maria Porras (University of Florida) co-founded #LatinX in Biomedical Engineering in February 2019, to leverage community as a means to increase representation of Latinos and promote diversity and inclusion in the Biomedical Engineering workforce. The platform leverages digital tools including Slack and Twitter and organizes webinars and virtual discussions to provide community mentoring and resources for Biomedical Engineers in academia. Since its inception, the #LatinXinBME community has provided mentoring to undergraduate students during academic admissions cycles, graduate students during fellowship and award applications, postdocs navigating the academic job market, and early career faculty members [10].

**#NativesInSTEM**
The #NativesInSTEM hashtag was first used in 2013 by Twitter user @native_engineer, a professional engineer who prefers to keep their identity anonymous; they were looking to foster an intersection of science and Native Peoples. Today, #NativeInSTEM is commonly used by a variety of Indigenous students, scientists and science organizations across institutions, countries and disciplines, as a means to find a digital community. Often, Indigenous students are the only Indigenous person in a department or field, as Indigenous students are less than 1% of undergraduate college demographics, and 0.6 percent of doctoral degree holders (2020) [17]. The #NativeinSTEM hashtag has empowered scholars to create community that spans into real life friendships for scholars that might otherwise feel invisible, and whom statistics often leave out because Indigenous populations are such a small part of the educational demographics [18]. These digital communities have also resulted into real life relationships and collaborations, such as the 2022 #WaterBack paper being written by a coalition of Indigenous scholars who first met through Twitter. An important note with the #NativeinSTEM movement, is that with over 572 federally recognized tribes, some Indigenous scholars will also use their specific tribal affiliation in place of the general #NativeinSTEM or #Indigenousscientist hashtags.

There has been tremendous energy and growth in #XinSTEM groups in a short time. We discussed how the dynamism of the movement can be expanded and leveraged to effect change, and how lessons learned can be translated to other components of the STEM ecosystem such as professional societies and funding institutions.

## LEVERAGING #XINSTEM TO BROADEN SOCIAL MEDIA STRATEGIES OF PROFESSIONAL SOCIETIES AND SCIENCE FUNDERS

Major scientific professional societies are actively looking for ways to engage with #XinSTEM social media movements. For example, the #BlackInPhysics team has engaged collaboratively with the National Society of Black Physicists (NSBP) group, while the #BlackInMath team has collaborated with the National Association of Mathematicians (NAM). We highlight a subset below:

**@SACNAS**

The Society for Advancement of Chicanos/Hispanics and Native Americans in Science ([SACNAS](#)) is creating new units within their existing infrastructure to integrate external social media efforts and to grow their own diverse social media communities.

SACNAS has a strong social media presence—not only through their @SACNAS national account and the hashtag #TrueDiversity, but also through individual chapter accounts. There are 133 chapters across the United States and U.S. territories, usually in colleges and universities, but also in industry and government.

Through their social media platforms (Twitter, Facebook, Instagram, LinkedIn, and YouTube), SACNAS reaches thousands of members and conference attendees every day. For example, the national @SACNAS Twitter account has 24.1K followers, and the SACNAS LinkedIn and Instagram accounts have ~5K followers each at the time of writing.

**1400 Degrees**

Private foundations are joining forces with #XinSTEM movements to use the directories of those communities to amplify and spotlight the voices of underrepresented groups in STEM. The Heising-Simons Foundation recently supported the #BlackinPhysics community with an award that funds their celebration week for 2021–2023.

At the same time, Heising-Simons also started [1400 Degrees](#), which is a directory dedicated to highlighting the achievements of transformative physicists and astronomers. The goal of the directory is to highlight women scientists and innovators and inspire all marginalized and non-conforming genders in order to promote gender equality.

Today, 1400 Degrees is seeking partnerships with various #XinSTEM groups to invite researchers into their database so their stories can be spotlighted on their platform. Note that the directory's name comes from 1400°C, the temperature at which glass begins to melt. By turning up the heat, 1400 Degrees hopes to break the glass ceiling for good.

**Burroughs Wellcome Fund (BWF)**

The Burroughs Wellcome Fund supports several #XinSTEM initiatives through both competitive and strategic ad-hoc award mechanisms. BWF interventions to promote diversity in STEM include the [BWF Postdoctoral Diversity Enrichment Program](#) (PDEP), which made its first awards in 2013 and has continued on an annual basis since then.

PDEP award recipients have become STEM leaders thanks to near-peer mentoring activities. One example is Promoting Engagement in Science for Underrepresented Ethnic and Racial Minorities (P.E.E.R). This partnership between Vanderbilt University, the University of Iowa, and Winston Salem State University was designed to establish inclusive, long-term, near-peer virtual mentoring within K–12 public schools.

Burroughs has also created a graduate diversity enrichment network (GDEN) consisting of both prior and current grantees. The network serves as a community of individuals who have not only received and benefited from diversity-based awards but who seek to "pay it forward" through their own engagement in diverse STEM activities. The hashtag #bwfpdep is used to drive these activities and cultivate a community of scholars.

## ACCELERATING THE PACE OF CHANGE REQUIRES RESOURCES

#XinSTEM movements nationwide are accelerating the representation of underrepresented groups in STEM. Unfortunately, many #XinSTEM initiatives lack the funding and resources they need to adequately scale up and plan for the long term. Most #XinSTEM programs are created with sweat equity and bootstrapped by crowdsourced funding (e.g., GoFundMe, Venmo, and CashApp campaigns. Such inconsistent funding methods are inadequate. There is an urgent need to create effective and *intentional* funding strategies that build sustainability for growing #XinSTEM movements.

**The problem with current funding strategies**

According to a survey commissioned by the Alfred P. Sloan Foundation on expenditures within STEM and presented by Dr. Lorelle Espinosa at our Salon, STEM higher education institutions received over $2 billion of private philanthropic investments from 2016–2020. However, just $123.9 million (5.8% of the total grants) went toward diversity, equity and inclusion programs (Figure 2) [19]. Surprisingly, the majority (44%) of the $2B investments were apportioned to a small group of 10 well-endowed, elite institutions, with nominal programs geared towards rectifying systemic barriers in STEM [20]. Such investment patterns only serve to uphold the current inequities within STEM and support small cohorts of students at elite organizations, rather than fostering systemic change. Financial reallocation of resources is required to provide resources for dynamic movements like #XinSTEM.

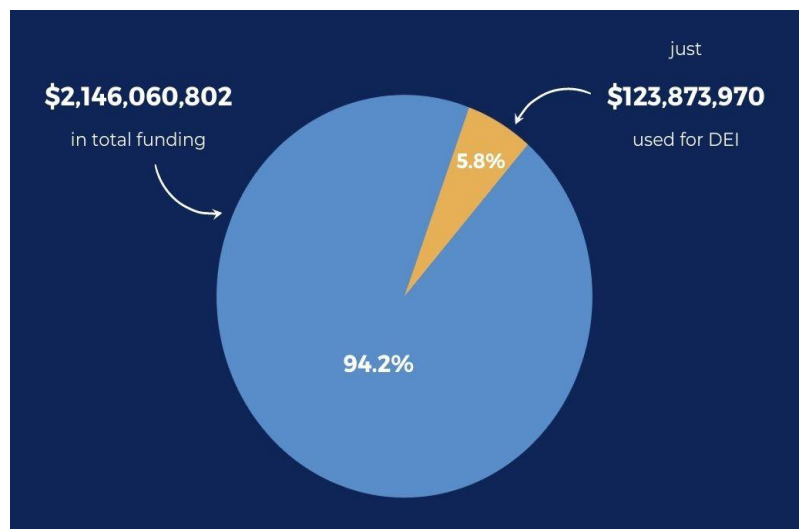

**Figure 2: Funding disparity in higher education institutions.** Funds for Diversity, Equity, and Inclusion (DEI) initiatives are a small slice (5.8%) of the total $2 billion USD allocation.

**Examples of funding programs that move the needle**

In 2021 the National Science Foundation (NSF) created the Mathematical and Physical Sciences Ascending Postdoctoral Research Fellowships (MPS-ASCEND), which supports postdoctoral MPS fellows in the U.S. and "provide[s] them with experience in research that will broaden perspectives, facilitate interdisciplinary interactions and help [broaden] participation within MPS fields."

Another timely NSF initiative is the [Leading Culture Change Through Professional Societies of Biology](#) (BIO-LEAPS) program, which is "designed to foster the necessary culture change within biology to move towards an equitable and inclusive culture that supports a diverse community of biologists that more fully reflects the demographic composition of the US population [21]."

New diversity, equity and inclusion-focused grant opportunities were also introduced recently, including the NIH Common Fund's [Faculty Institutional Recruitment for Sustainable Transformation](#) (FIRST)—a program supporting institutions that "enhance and maintain cultures of inclusive excellence in the biomedical research community [22]."

Recently, Burroughs Wellcome Fund increased the number of annual PDEP awards from 15 to 25 because of the success of the program. To date, 124 awards worth $7.5 million have been made. Three co-founders of the Black Microbiologists Association (the non-profit that grew from #BlackInMicro) are PDEP grant awardees: Dr. Nikea Pittman, a postdoctoral fellow at the University of North Carolina at Chapel Hill; Dr. Chelsey Spriggs, a postdoctoral fellow at the University of Michigan; and Dr. Ninecia Scott, a postdoctoral fellow at the University of Alabama at Birmingham.

## RECOMMENDATIONS FOR SUSTAINING #XINSTEM'S SUCCESS

#XinSTEM leaders, professional science societies members, decision-makers from academic institutions and funding bodies participated in the 2030STEM salon and repeatedly expressed the urgency for accelerating change in the STEM ecosystem. They described the exceptional opportunity at hand in which there is an outsized sensitive dependence on recognizing and reimagining STEM's initial conditions that can lead to bold, sustainable, and systemic changes for current and future STEM generations. To capitalize on this moment, we suggest the following recommendations:

**1. Create connections among #XinSTEM grassroots movements.**

Centralized coordination of funding support for #XinSTEM organizations would ensure broader access to vital opportunities throughout these underserved communities. Additionally, a centralized hub for data collection, reporting, and administrative support will be key to efficient funding allocation and distribution.

The time commitment required to support the momentum of a growing #XinSTEM movement is proportional to a full-time job. Many #XinSTEM groups rely on teams of volunteers to manage their workload. Becoming an independent entity, such as an LLC, can provide access to the funds needed to hire dedicated staff.

However, becoming an LLC brings on added responsibilities that may burden many early-career researchers. An umbrella organization to house several #XinSTEM movements would alleviate the burden of working *in silo*, and allow for a best-practice roadmap, drawn from the experiences of successful initial groups.

**2. Make diversity, equity and inclusion a major criterion for new funding.**

Philanthropy and government granting agencies are pivoting to focus on systemic change strategies in their funding. Funding bodies should use their awards dollars like "carrots and sticks."

"Carrots," such as the NSF-BioLEAPS program, are a method for funders to employ grantmaking with a systemic-change lens. As suggested by the Sloan survey, this type of funding should prioritize minority-serving institutions and those "doing the work" like #XinSTEM founders.

"Sticks" can be used to hold awardees accountable. Funders should withhold future awards or allocations until certain programs are initiated that will increase full STEM inclusion. For example, NSF can more stringently assess the Broader Impacts projects in research awards to ensure that their activities are increasing inclusion over time.

**3. Implement service and scholarship metrics.**

Make the efforts of #XinSTEM founders a part of their scholarship and service metrics for the tenure and/or promotion process at their institutions. For example, building professional development into the grantmaking and/or awards process could greatly improve the ability of awardees to meet and exceed key metrics for career advancement.

Recognizing diversity, equity and inclusion work as a form of STEM community-building can also ignite institutional culture change. Adopting strategies that showcase service and scholarship on a level playing field, as is being attempted with the revised [Résumé for Researchers](#) models, may help achieve these goals.

**4. Start local chapters at #XinSTEM founder institutions.**

One of the "low-hanging fruit" ways to support #XinSTEM communities is to simply support the establishment of local chapters at #XinSTEM founder institutions and across professional societies.

The momentous achievements of #XinSTEM founders should be celebrated and recognized by the institutions they attend and the STEM professional societies they represent. Official recognition from their home institutions and professional societies, along with the financial support necessary to create local chapters, could alleviate much of the burden in adequately funding #XinSTEM communities to ensure their longevity.

For example, #LatinxinMarineScience did not seek to become an LLC because it was being run by four graduate students in their limited spare time. Efforts to seek funding were unsuccessful, and #LatinxinMarineScience was only operational for one year. Operational support from the local institution where #LatinXinMarineSciences was founded or professional societies that support marine sciences may have made an impact to keep the movement active.

Strategies such as those found within the National Science Foundation BIO-LEAPs initiative could be amplified to accelerate engagement of #XinSTEM movements and professional science organizations.

**5. Collaboratively build #XinSTEM databases**

#XinSTEM movements, together with platforms and organizations such as 1400 degrees and GDEN, are creating comprehensive databases of scientists that include underrepresented groups, such as community pages of #BlackinNeuro, #BlackinMicro, #BlackinChem, and #BlackinCardio [23-26]. More deliberate collaborations to build databases of underrepresented groups in STEM that amplify their exposure will increase access to critical opportunities for career development and increase STEM-identity in next generation students.

Inclusive STEM databases are an invaluable resource. They can be accessed and mined for activities such as serving as grant and manuscript reviewers, nominations for academy memberships, considered for prestigious awards, keynotes for conferences, and most importantly, provide a pathway to fresh voices with different and valuable perspectives.

## CONCLUSION

Movements like #XinSTEM have enabled researchers, students, and aspiring scientists to connect, engage, and amplify underrepresented voices in their field. Since 2020, the ability of #XinSTEM (and other social-media-based grassroots movements) to foster real change continues to grow. While it will require collaboration between scientists, educational institutions, and granting agencies and organizations, the precedent set by #XinSTEM and its prodigious expansion will further inclusion and representation within the STEM ecosystem.

One of the most profound measures of success is increased representation of Black, Latino/a/x, and Indigenous individuals and other underrepresented groups in STEM. In order to achieve full inclusion across STEM fields, systematic changes must be made. #XinSTEM leaders have shown us how to generate excitement and create welcoming spaces. The challenge on everyone in the STEM ecosystem is to find ways to leverage these connections, synthesize best practices, create sustainable funding structures that support inclusion at systemic levels and grassroots efforts like #XinSTEM, to ensure the next generation of STEM practitioners are a full demographic representation to tackle the global science and technology challenges ahead.

**ACKNOWLEDGMENTS**

2030STEM Inc. gratefully acknowledges funding from the Alfred P. Sloan Foundation (G-2021-16977) and for their inspirational support in our planning year and for our Salon series. 2030STEM also acknowledges all participants of the #Change Salon for their thoughtful insight, visionary contributions, and dedication to building a STEM ecosystem that works for all. Holford's work was also supported by National Science Foundation Award (DRL # 2048544).